\documentstyle[epsfig,12pt]{article}

\newcommand{\be}{\begin{equation}}
\newcommand{\ee}{\end{equation}}
\newcommand{\sbbox}[1]{\mbox{\scriptsize\boldmath $#1$}}

\begin{document}

\title{
Continuum limit of finite temperature $\lambda\phi^4_3$ from 
lattice Monte Carlo. 
}

\author{G. Bimonte$^{a,b}$, D. I\~niguez$^{a}$, A. Taranc\'on$^{a}$ and 
C.L. Ullod$^{a}$}

\maketitle

\begin{center}
{\it a)  Departamento de F\'{\i}sica Te\'orica, Facultad de Ciencias,\\
Universidad de Zaragoza, 50009 Zaragoza, SPAIN \\
\small e-mail: \tt david, tarancon, clu@sol.unizar.es} \\
{\it b)  Dipartimento di Scienze Fisiche, Universit\'a di Napoli,\\
Mostra d'Oltremare, Pad.19, I-80125, Napoli, Italy; \\
INFN, Sezione di Napoli, Napoli, ITALY.\\
\small e-mail: \tt bimonte@napoli.infn.it } \\
\end{center}

\begin{abstract}
The $\phi^4_3$ model at finite temperature is simulated on the lattice.
For fixed $N_t$ we compute the transition line for $N_s \to \infty$ by
means of Finite Size Scaling techniques. 
The crossings of a Renormalization Group trajectory  with the
transition lines of increasing $N_t$ give a well defined limit for 
the critical temperature in the continuum.
By considering different RG trajectories, we compute 
$T^c/g$ as a function of the renormalized parameters.
\end{abstract}

\vfill
DFTUZ/96/21  \space\space DSF 96/50  \space\space  hep-lat/9610038 

\newpage

\section{Introduction}

Finite temperature phase transitions in Relativistic Quantum Field Theories
have been the subject of an intense study since the time of
the pioneering work of  Kirzhnitz and Linde \cite{kir},
Weinberg \cite{wei} and Dolan and Jackiw \cite{dol},
which revealed that, in general, spontaneously broken symmetries get
restored above a certain critical temperature $T^c$
\footnote{There exist models, however, which do not seem to follow
the general rule and so exhibit symmetry breaking 
at arbitrarily large temperatures \cite{wei,dva}.
Whether this is
a genuine effect or an artifact of (lowest order) perturbation theory is
still a subject of discussion \cite{fuj2,bim,fuj3}. }. 
While the nature of this phenomenon is by now
well understood, an accurate description of the 
transition, including sometimes its order,
as well as a precise determination of
the critical temperature can be difficult because of the appearance,
at the transition point, of
infrared divergencies which cause the breakdown of
ordinary perturbation theory \cite{dol}. 
Even though
new analytic non-perturbative methods, allegedly immune of 
infrared problems, have recently been developed and have been 
used to attack this sort of problems \cite{tet}, 
the failure of the standard analytic methods has 
motivated some authors to study these phase transitions 
on computers and these days massive simulations of
the electroweak phase transition are being performed
(see for example \cite{far}).

In this paper we report on the results of a lattice Monte Carlo
simulation of the $\lambda \phi^4_3$ theory 
at finite temperature. There are definite reasons to be interested in
this simple system. First of all it is 
believed that the 
$\phi^4_3$ theory admits a non-trivial continuum
limit (at $T=0$) \cite{coo,fre,dom,co2,wes}, contrary to $\phi^4_4$. 
Moreover, $\phi^4_3$ is superrenormalizable \cite{zinn},
so that primitive ultraviolet divergences occur only in a finite
number of diagrams, and this allows one to exactly compute its
$\beta$ functions. The model is expected to undergo a second
order phase transition at a finite critical temperature
but despite its simple ultraviolet properties,
its behavior near the transition is particularly
hard to explore, because infrared divergencies get worsened
in three dimensions and as a matter of fact there is no reliable way
of computing the critical temperature in perturbation theory,
not even to lowest order \cite{fuj}.
Finite Temperature Renormalization Group methods may nevertheless
be used to get estimates of the critical temperature
\cite{ein}.

It appears natural to see if better results can be obtained 
from the lattice. In this respect, there exist
theorems proving that on the lattice spontaneously broken symmetries 
get always restored at sufficiently high temperatures
(in an arbitrary number of dimensions)
\cite{kin,bim}. Unfortunatly, these
theorems only provide upper bounds on the critical temperature
(for finite lattice spacings),
which cannot be used to determine the critical temperature
in the continuum limit,
not being strong enough to even guarantee that the critical temperature
remains finite when the lattice spacing is taken to zero. 

Of the existing lattice studies of the $\phi^4_3$-theory at 
finite temperature, some combine
approximate analytical treatments with numerical work
to get estimates of the critical temperature in the continuum
limit  \cite{fuj3,li}, while others use a Monte Carlo simulation 
\cite{ama}. This paper represents a significant
improvement and extension of these results. 
We have carried out a Monte Carlo simulation on large lattices 
with massive statistics, making a detailed study of 
finite size effects and of the continuum limit.

Within the path integral formalism, finite temperature is introduced
by compactifying the euclidean time direction, temperature being the
inverse of the temporal extension \cite{kap}. When doing simulations on the
lattice, this goal is achieved by using asymmetric lattices 
$N_t\times{N_s}^2$
shorter in the temporal direction. However, in order to avoid finite size 
effects and approach the continuum limit, these steps should be followed:
\begin{enumerate}
\item{} For a finite $N_t$ take the limit $N_s\to\infty$.
\item{} Take the lattice spacing $a$ to 0 and $N_t$ to $\infty$.
\end{enumerate}

Our aim is to measure the critical temperature for which symmetry restoration
arises. For fixed $N_t$ we have computed, 
for several values of $N_s$, the transition
line that separates the broken and symmetric phases. By means of Finite Size
Scaling (FSS) techniques we have obtained the line for $N_s\to\infty$. 
We have done this for
several values of $N_t$. When increasing $N_t$, the transition lines
approach that of zero temperature.
These critical lines have been used to measure $T^c$ in the
following way. A possible continuum theory is defined
by a ``curve of constant physics'' (CCP): it is parametrised by
the lattice spacing $a$ and, as $a\rightarrow 0$, it  
approaches the critical line of the symmetric lattice.
While doing so, it intersects the critical lines of the lattices with
finite $N_t$. Each intersection point gives a critical temperature, 
$T_{N_t}=\frac{1}{N_t a}$, and $T^c$ is just the limit (if it exists)
of all these temperatures.

Due to the super-renormalizability of the
model, the asymptotic form of the CCP's that approach the gaussian fixed point
can be computed exactly in perturbation theory. We have
considered several CCP's and have accepted only those 
for which the correlation length, measured numerically, has shown a 
satisfactory scaling.
For these trajectories we have found
a well defined limit for the critical temperatures and we have been able to
verify the relation predicted in \cite{ein}
between $T^c$ and the renormalized parameters
of the continuum model (defined using dimensional regularization
in the Minimal Subtraction (MS) scheme)
\be
T^c \log \left( \frac{4 \pi k T^c}{g} \right)
= - \frac{8 \pi m^2_{MS}
(T^c)}{g}~,
\ee
where $m_{MS}(T^c)$ is the value of the running mass
at the scale $T^c$, $g$ is the coupling constant and $k$
is a factor not calculable perturbatively. We have obtained
an estimate of $k=2.1(2)$.

The plan of the paper is as follows. In Sec.2 we present the
model. Sec.3  defines the curves of constant physics. 
In Sec.4 we discuss the effects of a finite temperature and 
explain how to measure its critical value. Sec.5 describes the 
details of the Monte Carlo 
simulation and the observables that have been measured. The results are
presented in Sec.6. Finally, Sec.7 is devoted to the conclusions.

\section{The $\lambda\phi^4_3$ model}

We consider the theory for a real scalar field in 3 euclidean 
dimensions, described by the bare (euclidean) action:
\begin{equation}
S=\int d^3 x \left[\frac{1}{2}
(\partial_i\Phi_B)^2 + \frac{1}{2}m^2_B
\Phi_B^2 + \frac{g_B}{4!} \Phi^4_B\right]~~.\label{bact1}
\end{equation}

When regularized on an infinite three-dimensional cubic 
lattice of points
$\Omega$ with lattice spacing $a$, the above action is replaced
by the discretized expression:
\be
S_L=\sum_{x \in \Omega} a^3 \left\{ \frac{1}{2} \sum_{\mu}
(\Delta^{(a)}_{\mu}
 \Phi_L)^2(x)
+\frac{1}{2}m^2_B
\Phi_L^2(x) + \frac{g_B}{4!} \Phi^4_L (x)\right\}~~\label{lact1}
\ee
where $\Delta^{(a)}_{\mu}$ is the lattice derivative operator in the
direction $\mu$
$$
\Delta^{(a)}_{\mu} \Phi_L(x) \equiv 
\frac{\Phi_L(x+\hat{\mu}a)-\Phi_L (x)}{a}~~.
$$

It is convenient to measure all quantities in the action (\ref{lact1})
in units of the lattice spacing. We thus define:
\be
\phi_L(x) \equiv a ^{1/2} \Phi_L(x),~~~u \equiv a g_B,~~~
r \equiv a^2 m^2_B~~.\label{par}
\ee

In terms of the rescaled quantities the action now reads:
\be
S_L=\sum_{x \in \Omega} \left\{ \frac{1}{2} \sum_{\mu}
(\Delta_{\mu}^{(1)} \phi_L)^2(x)
+\frac{1}{2} r
\phi_L^2(x) + \frac{u}{4!} \phi_L^4 (x)\right\}~~.\label{lact}
\ee

In order to conform with the customary lattice notation, we perform a
further redefinition of the field and parameters in (\ref{lact}):
\be
\phi_L(x) \equiv {\sqrt \kappa} 
\phi_{\sbbox{r}},~~~u=\frac{24 \lambda}
{\kappa^2},~~~
r=2\frac{1-2\lambda -3 \kappa}{\kappa}~~
\ee
and thus write the action in our final form: 
\begin{equation}
S_{L}=  \sum_{\sbbox{r} \in {\bf Z^3}}\left\{ 
      -\kappa  \sum_{\mu}\phi_{\sbbox{r}} \phi_{\sbbox{r}+
\hat{\sbbox{\mu}}}+
       \lambda(\phi^2_{\sbbox{r}}-1)^2+\phi^2_{\sbbox{r}}\right\}\ 
\label{fin}.
\end{equation}

This model has been intensely studied at
zero temperature, using renormalization
group and high temperature expansion techniques \cite{dom}, and
also numerically \cite{coo,fre,co2,wes}, and its features are well known.  
As it is apparent from the action, it has a $Z_2$ symmetry which 
can be spontaneously broken, if the double-well potential is
deep enough. In the 
parameter space $(\kappa,\lambda)$ there exist two fixed points: 
one is the gaussian fixed point
$(\kappa=1/3, \lambda=0)$ while the other is
in the same universality class as that of the 3d Ising model. 
The gaussian fixed point has a null attraction
domain and the whole transition line (except just the origin) possesses the
Ising critical exponents.

\section{The Curves of Constant Physics}

In this section we define the CCP's which define the
continuum limit of our lattice model. They are Renormalization Group
trajectories of the theory at zero temperature.

Suppose we start from a symmetric $N^3$ lattice with spacing $a_0$ and
bare parameters $m^2_B,~g_B$. 
In order to take the continuum limit, we have to vary $a$ in
the interval $[a_0,0)$ and simultaneously adjust the values
of $m^2_B,~g_B$, in such a way that physical observables
approach a constant value when $a\rightarrow 0$. 
If there were no need for renormalization, this would be 
a trivial task: it would suffice to hold $m^2_B$ and $g_B$ constant
in eq.(\ref{par}) and this would tell us how to adjust the 
dimensionless lattice
parameters $u$ and $r$ with $a$ in the approach to the continuum.
Unfortunatly, things are not so easy and the above procedure
would fail, for example, to produce in the continuum a finite
physical mass for the lightest boson. 
Because of the interaction, it is necessary to give
the lattice parameters a non trivial dependence on $a$ in order
for the observables, like the masses, to approach a finite limit
for $a\rightarrow 0$ and we call ``curves of constant physics''
(CCP's) any  dependence of the lattice 
parameters $(u,r)$ (or equivalently $(\kappa,\lambda)$) on $a$
which achieves this goal.

In principle, a possible way of defining 
the CCP's would be by requiring that two observables 
$({\cal O}_1, {\cal O}_2)$, constructed out of the
Green's functions of the theory (for example the physical mass
of the boson and the coupling constant at some fixed momentum scale), 
keep a constant value in physical units when $a$ is varied. 
The CCP will depend of course on the choice of 
$({\cal O}_1, {\cal O}_2)$,
but for $a$ sufficiently small they will all coincide and define
the same continuum physics, possibly parametrized in
different manners. 

As from a numerical point of view this strategy is very hard to
implement, 
we have utilized a different procedure to
construct our CCP, based on
the fact that our model is perturbatively super-renormalizable.
Due to this property, the perturbative expansion of the
theory contains only a finite number of 
ultraviolet primitively divergent diagrams \cite{zinn}. Concretely,
depending on the regularization
scheme, such divergencies can occur only in the 
one-loop ``tadpole'' diagram and in the two-loop ``sunset''
diagram, and both can be reabsorbed via a redefinition
of the mass term. As a consequence, no infinite
renormalization of the field and of the coupling constant
are needed and we can thus identify the bare field $\Phi_B$
and the bare coupling $g_B$  in (\ref{bact1}) with the renormalized ones,
$\Phi_R$ and $g_R$, respectively. We can then rewrite the action
(\ref{bact1}) as:
\be
S=\int d^3 x \left[\frac{1}{2}
(\partial_i\Phi_R)^2 + 
\frac{1}{2}m^2_B
\Phi_R^2 + \frac{g_R}{4!} \Phi^4_R\right]~~.\label{bact}
\ee
(From now on we shall omit writing the subscript $R$ of all
renormalised quantities, for simplicity.)
In this expression, the only divergent quantity is thus $m^2_B$.
We split it as: 
\be
m^2_B=m^2 + \delta m^2
\ee
where $m^2$ represents the tree level mass term
and $\delta m^2$ is a mass-counterterm which diverges
in the limit $a \rightarrow 0$. The expression of $\delta m^2$
depends of course on the renormalization scheme. Its
computation becomes especially easy in the
lattice analogue of the minimal subtraction scheme, because
all is needed then are the divergent parts of
the lattice tadpole and sunset diagrams. All the formulae needed
for this computation can be found, for example, 
in \cite{far} and here we just quote the answer (for the limit
$N \rightarrow \infty$):
\be
m^2_B= m^2(\mu)  - \frac{\Sigma}{8 \pi a} g-
\frac{1}{96 \pi^2}g^2 \ln (a \mu)~,\label{mass}
\ee
where $\mu$ is an arbitrary mass scale and $\Sigma=3.1759114$. 

With the above expression for the bare mass, 
the perturbative expansion of the Green's functions
of the theory is finite \underline{to~all~orders} and thus all
the observables that can be constructed out of them
have a well defined limit for $a\rightarrow 0$. This
suggests that we define as CCP those
that simply have a fixed value of $m^2$ (at some fixed scale $\mu$) 
and of $g$. A similar approach was used and shown to work properly for 2d
$\lambda \phi^4$ \cite{ciria}.  

Using eq. (\ref{mass}) and the definition of the
lattice parameters eq.(\ref{par}), it is easy to see that the
CCP, starting from some
initial point $a_0$, $r_0$ (at the scale $\mu_0$), 
$u_0$, takes then the form
$$
u(s)=s u_0~,
$$
\be
r(s)=s^2\left( r_0 + \frac{\Sigma}{8 \pi} u_0
\left(1 - \frac{1}{s}\right) - \frac{1}{96 \pi^2}u_0^2 \ln s\right)~,
\label{rge}
\ee
where $s=a/a_0$. Notice that, according to these formulae,
the CCP's approach the gaussian fixed point, as it must
be having been computed within perturbation theory.

The parameters $m^2$ and $g$ just represent 
renormalised quantities and do
not constitute a pair of observables that can be
measured directly on the lattice.
Consider instead one such observable, for
example the physical mass of the lightest 
particle, $m_{phys}$, which is simply related to 
the correlation length at large separations $\xi_{phys}$, 
since $m_{phys}=\xi_{phys}^{-1}$. Now, when computed
in perturbation theory starting from (\ref{bact}), with $m^2_B$ given
by (\ref{mass}), $m_{phys}^2$ becomes a function of
$m^2,~g$ and $a$, admitting a finite limit for $a\rightarrow 0$.
On dimensional grounds we can write this function as
$$
m_{phys}^2=m^2 F(g/m,am).
$$

When the continuum limit is sufficiently near, 
the dependence on $a$ of the r.h.s. becomes
negligible, $m_{phys}$ and thus
the correlation length $\xi_{phys}$ become independent
on $a$ and this implies that the lattice correlation length
$\xi= \xi_{phys}/a$ becomes proportional to $1/a$.

In terms of the parameter $s$ introduced earlier, the
region
of continuum physics, along the CCP, is that for which the
lattice correlation length scales according to 
\be
\xi(s) \propto \frac{1}{s}~. \label{scal}
\ee

Moreover, one should always have $\xi>1$ in order for the lattice
details to be irrelevant and $\xi<N$ in order for finite size effects
to be negligible.

\section{Finite temperature}

We briefly discuss the effects of a finite 
temperature on our system and how to measure $T^c$ on the lattice.

Let us suppose that the system is in the ordered
phase at $T=0$ and imagine heating it up. 
The intuitive picture, valid in general, is that disorder increases,
as the temperature is raised, 
until one reaches a critical value
$T^c$ above which no order is possible any more
and the full symmetry of the lagrangian is restored.
This simple picture is confirmed by analytical treatments
in four space time dimensions \cite{wei,dol}. 

In three dimensions things are less simple. Consider
our $\phi^4_3$ model: while the
existence of the symmetry restoring phase transition at a
finite temperature seems sure,
a reliable perturbative computation of the critical temperature,
using standard techniques, is problematic 
due to the presence of severe
infrared divergencies. For example, when taking  
a high temperature expansion of the one-loop effective
potential one finds that the leading field-dependent term
becomes complex for small fields,
and this leads to a physically unacceptable complex
value for the critical temperature. The situation does
not improve upon resumming the so called daisy and super-daisy diagrams
\cite{dol}: the
resulting gap equation determining the thermal mass
becomes meaningless at the critical temperature \cite{fuj}.

To our knowledge, the best analytical estimate of $T^c$ for the
$\phi^4_3$ model
is the one recently obtained in \cite{ein}
using renormalization group methods. It is given in terms
of an implicit equation for $T^c$:
\be
\frac{T^c}{g} \log \left( \frac{4 \pi k T^c}{g} \right)
= - \frac{8 \pi m^2_{MS}
(T^c)}{g^2}~,\label{ein}
\ee
where $m^2_{MS}(\mu)$ is the running mass
\be
m^2_{MS}(\mu)=m^2_{MS}(\mu_1)+\frac{g^2}{96 \pi^2}\log\left(
\frac{\mu}{\mu_1}\right)~.\label{run}
\ee
According to \cite{ein}, the uncertainty on the value of $T^c$  
due to the breakdown of perturbation theory in the vicinity
of the phase transition only affects the value of the number
$k$ in eq.(\ref{ein}). It cannot be computed analytically
but is expected to be of order one and, as will be shown 
later, we have been able to measure it quite accurately.

The difficulties on the analytical side motivated
us to study the phase transition of $\phi^4_3$ on the lattice.
In order to simulate a finite physical temperature 
we work on an 
$N_t\times{N_s}^2$ lattice. For every
$N_t$ considered, we have performed a series of simulations
for several values of $N_s$, 
and using FSS techniques we have obtained the transition line in
the thermodynamic limit
$N_t\times {\infty}^2$. Repeating this analysis for different 
values of $N_t$ we have obtained several transition lines. 
For decreasing $N_t$ the critical lines
shift deeper inside the broken region of the 
symmetric lattice, though all of them meet at the gaussian fixed point 
($\kappa=1/3,\lambda=0$) whenever $N_t>1$. 

Consider now a CCP, call it $\gamma$,
lying in the ordered region of the $T=0$ theory and let $P$ be its
starting point. Since
the critical lines for finite $N_t$ 
approach, in the limit $N_t \rightarrow \infty$, that of the $T=0$ theory,
$P$ will also lie in the ordered regions of the lattices
with finite $N_t$, for $N_t$ large enough (how large it depends
on $P$, of course). Choose now one such $N_t$, and
imagine moving along the CCP. In this process $a$ diminishes
and thus the physical temperature associated with the $N_t$ chosen 
grows up since $T=1/N_t a$. At some
point our CCP will cross the transition line for this value of $N_t$ 
and symmetry will be restored. If $a_{N_t}$ is the
value of the lattice spacing along the CCP at the crossing point, we have
an estimate of the critical temperature:
\be
T_{N_t}
=\frac{1}{N_t a_{N_t}}   \label{est}
\ee

In order to get the continuum critical temperature we have to
let $N_t\rightarrow \infty$.
When $N_t$ is increased, the corresponding critical line
gets closer to that of the $T=0$ theory and
$\gamma$ will intersect it for a smaller value of $a$. The temperatures
$T_{N_t}$ are thus expected to approach a limit, 
which is the desired continuum critical temperature
$T^c$ for the continuum theory associated with
$\gamma$. 

In the absence of an external input which fixes the energy scale,
we can assign a value $a=a_0$ to an arbitrary point on the CCP 
and then the value of $a$ at any other point on the trajectory is given by 
equations (\ref{rge}). In this way we can compare the temperatures 
$T_{N_t}$ corresponding to different values of $N_t$ 
for a given CCP, but it is not possible
to put together the values from distinct CCP's due to the 
arbitrariness in the scale. To avoid this problem, one can measure 
adimensional quantities, obtained for instance from the quotient of two
observables with the same dimensions. Of this sort are the
quantities $T^c/g$ and
${m^2_{MS}}(T^c)/ g^2$ entering equation (\ref{ein}).
 
The analysis of \cite{ein} was carried in the continuum,
using dimensional regularization and minimal subtraction,
while we are using the lattice regularization. So, 
in order to compare the results of our simulations
with eq.(\ref{ein}), we have to first relate the continuum 
renormalized parameters appearing in eq.(\ref{ein}) to ours.
In view of what we said earlier
about the ultraviolet properties of the model, the only
non trivial relation that needs to be computed 
is the one among the mass parameters of the two schemes.
Using the formulae of \cite{far} for the relevant two-loops 
diagrams contributing
to the self-energy, it is not hard to check that:
\be
m^2_{MS}(\mu)=m^2(\mu)-\frac{g^2}{96 \pi^2}(\log 6 + \zeta)
+ o(g^3)~,\label{ms}
\ee
where $\zeta=0.09$ and $m^2(\mu)$ is the lattice renormalized
mass defined in eq.(\ref{mass}). 
Upon using eqs.(\ref{mass}), (\ref{ms}) and (\ref{est}), we can now
relate the adimensional quantities
$T^c/g$ and ${m^2_{MS}}(T^c)/ g^2$ 
to the simulation parameters according to the
following equations:
\be
\frac{{m^2_{MS}}(T^c)}{g^2}=\lim_{N_t \rightarrow \infty}
\frac{1}{u^2}\left( r + \frac{\Sigma}{8 \pi} u
- \frac{1}{96 \pi
^
2}u^2 (\log N_t + \log 6 + \zeta)\right),
\ee
\be
\frac{T^c}{g}=\lim_{N_t \rightarrow \infty}
\frac{1}{N_t u}
\ee
where $u$ and $r$ are evaluated at the points in which
the CCP intercepts the critical line of the $N_t$-lattice. In
our simulation we have measured the above quantities for various 
CCP's, and this allowed us to check the validity of eq.(\ref{ein})
and measure, at the same time, the value of the unknown
parameter $k$.

\section{Monte Carlo simulation and observables}

In the course of the simulations, most of our efforts went into
the computation of the critical lines for different
values of $N_t$. On the other hand, we have made measurements 
also on the symmetric lattice in order to identify the scaling region
of our perturbative CCP's.

We have used a Metropolis algorithm 
combined with a Wolff single cluster method 
\cite{wolff} (the latter
updates the sign of the field), taking a ratio of 20 clusters 
every 3 Metropolis iterations or
10 clusters every 2 Metropolis. These values are inside the region of ratios
we have found nearly optimal for decorrelation versus CPU time close to
the critical lines. 
When, in the intent of checking the CCP's, we have measured the
correlation lengths on the symmetric lattices, 
the simulations have been performed in the ordered region of the phase
diagram and there the clustering is not so effective. However these simulation 
points did not lie very deep in the broken phase and then we have used 
this method as well.

The most intensive simulations have been those relative to the
asymmetric lattices.
We have used four different values of $N_t$ $(2,3,4,6)$ and for 
each of them 
we have computed the critical value of $\kappa$ for four 
values of $\lambda$ 
($0.0005$, $0.001$, $0.002$, $0.003$). For each pair $(N_t,\lambda)$, 
we have considered
lattices of $N_s=12,16,24,32$, performing on each of them a single
simulation  for one value of $\kappa$. This value
has been chosen near the critical one,
by means of an hysteresis or by
extrapolating from the values found for the previous lattices.
Afterwards,
we have used the spectral density 
method (SDM) \cite{ferr} to extrapolate the 
measurements to other values of $\kappa$ in the vicinity of 
the simulation point.
Finally, the value of $\kappa$-critical in the
limit $N_s\to\infty$ has been computed by examining
the crossings of the Binder cumulants 
relative to lattices with different values of $N_s$,
as explained below.

At each of the points and lattices that have been simulated,
the number of iterations made has been approximately
$1,500,000\times(3~Metropolis+20~Clusters)$ or 
$2,500,000\times(2~Metropolis+10~Clusters)$, for a total CPU time equivalent 
to nearly one year of workstation. 
Most of the simulations have
been carried out with iterations consisting 
of $3~Metropolis+20~Clusters$ and performing
measurements only at intervals of 10 iterations. 
The errors in the estimation of the observables have been calculated 
with the jackknife method.

\begin{figure}[t!]
\epsfig{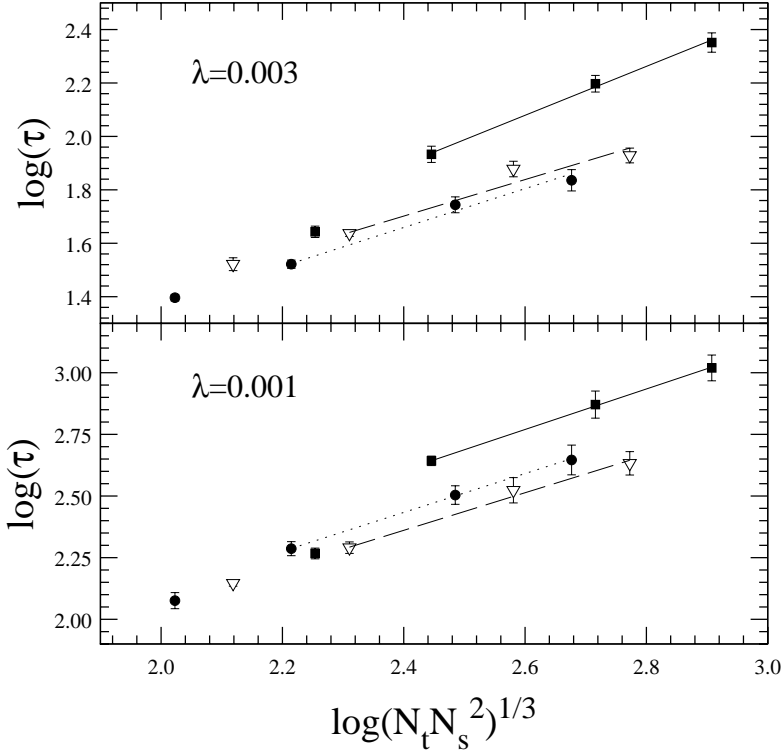}
\caption{ Logarithms of the autocorrelation times $\tau$ as a function of
$\log(N_t N_s^2)^{1/3}$, for $N_t=3$ (full circles), $N_t=4$ (empty triangles)
and $N_t=6$ (full squares) and $N_s=12, 16, 24, 32$. 
The straight lines are the fits for the
dynamical exponents $z$ according to eq.(\ref{dexp}). The points
not covered by the fits correspond to $N_s=12$ and have not
been used for the computation of the fits.} 
\label{TAU}
\end{figure}

We have measured the integrated autocorrelation times $\tau$. 
The values obtained from the $\kappa$-energy, $\lambda$-energy and 
magnetisation are totally compatible. 
They oscillate, in units of $10\times(3~Metropolis+20~Clusters)$, 
approximately from $\tau\simeq3$ in the best
case (largest $\lambda$ and smallest lattice) to $\tau\simeq38$ in the worst
(smallest $\lambda$ and largest lattice). 
In particular we have estimated the dynamical exponent $z$ defined by
\be
\tau \sim {(N_t\times{N_s}^2)}^{z/3}   \label{dexp}
\ee
with fixed $N_t$, obtaining $z=0.8(1)$. This
estimate is not totally rigorous because the autocorrelation times have 
been measured at the simulation points which do not coincide exactly with 
the 'critical' ones (but they do it approximately). 
However it should be basically correct because, as can be seen
in Figure \ref{TAU},
a power law with a rather well
defined exponent is observed
for the largest lattices. Figure 1 shows how $\tau$ scales 
with $N_s$, for fixed $N_t=3,4,6$
at $\lambda=0.001$ and $\lambda=0.003$.
The fits have been computed using only the data relative to
$N_s=16,24,32$, but for completeness in the figure we have displayed 
also the
points corresponding to $N_s=12$.
They give an exponent $z$ oscillating between $0.7$ and $0.9$ approximately.   
Metropolis alone would give a value $z=2$ while the Wolff algorithm, in spin 
systems where can it be used by itself, has $z=0$ because at the critical
point the clusters can have any size. In our case the clustering is 
only able to 
change the sign of the fields, but its non-locality contributes to 
diminish significantly the dynamical Metropolis exponent.  

For the sake of completeness, we have also determined the approximate 
position of the critical line at zero temperature. In this case no 
extrapolation to the thermodynamical limit has been made, and we
have simulated only a $16^3$ lattice. 

In order to estimate the critical points, we have used the Binder
cumulant defined as:
\be
U_{N_s}(\kappa)=\frac{3}{2}-\frac{\langle M^4 \rangle}{2\langle M^2\rangle^2}
\ee
where
\be
M=\vert (1/V) \sum_{\sbbox{r}} \phi_{\sbbox{r}} \vert
\ee
is the magnetisation or expectation value of the field, $V=N_t \times
N_s^2$ being the volume.

At several points on the CCP's, we have measured the second moment correlation
length on the symmetric lattice $16^3$, in order to check the scaling.
It is defined as \cite{coo,sokal}
\be
\xi=\left( \frac{1}{2d}\frac{\sum_{\sbbox{x}}{\vert x \vert}^2 G(x)}
                       {\sum_{\sbbox{x}}G(x)} \right) ^{1/2},
\ee
where $G(x)$ is the correlation function.
Usually this quantity is measured in the disordered phase, while
we need evaluate it in the broken phase. There $\langle M \rangle \ne 0$ 
and then we have to use the connected correlation function

\be
G(x,y)=G(x-y)=\langle \phi_x \phi_y \rangle - 
        \langle \phi_x \rangle \langle \phi_y
\rangle.
\ee

 Using the lattice symmetries, $\xi$ can be expressed as

\begin{equation}
\xi =\left(\frac{ \frac{\chi}{F} - 1}{4sin^2\frac{\pi}{N}}\right)^{1/2}
\end{equation}
where $N=N_s=N_t$ is the lattice size, $\chi$ is the susceptibility

\begin{equation}
\chi=\sum_{\sbbox{x}}G(x)=V( \langle M
^2 \rangle - {\langle M \rangle}^2)
\end{equation}
and $F$ is the analogous quantity at the smallest nonzero momentum
(having two null components and one equal to $\pm2\pi/N$) 

\begin{equation}
F=\sum_{\sbbox{x}}G(x)e^{i p \cdot x}=
\frac{1}{3V} \langle
{\vert \sum_{\sbbox{r}} \phi_{\sbbox{r}} e^{2\pi i r_1/L} \vert}^2
+{\vert \sum_{\sbbox{r}} \phi_{\sbbox{r}} e^{2\pi i r_2/L} \vert}^2
+{\vert \sum_{\sbbox{r}} \phi_{\sbbox{r}} e^{2\pi i r_3/L} \vert}^2
\rangle
\end{equation}
$r_j$ being the components of $r$ (indeed, the fact of being in the 
ordered phase is irrelevant as far as $F$ is concerned, 
because the disconnected part cancels out from the sum 
due to the Fourier factor).

\section{Results}

\subsection{Phase diagram}

\begin{figure}[t!]
\epsfig{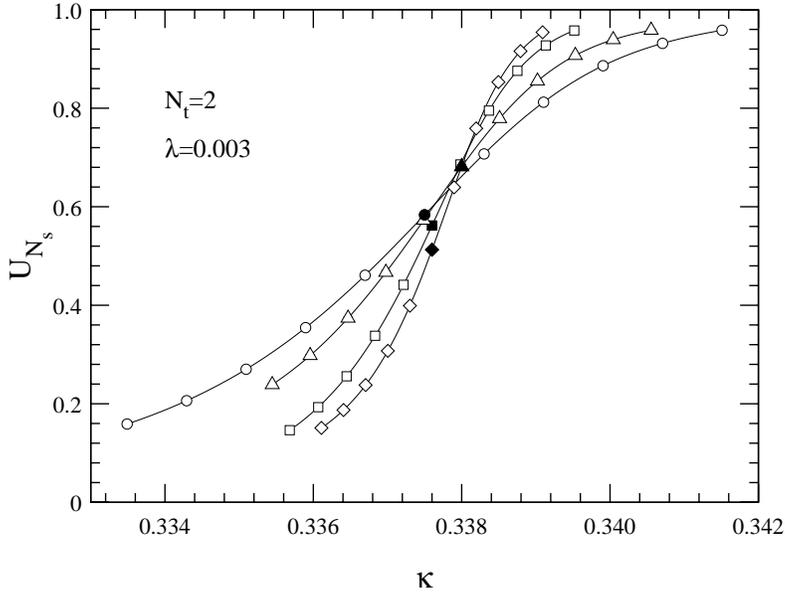}
\caption{The Binder cumulants $U_{N_s}$ as a function of the
hopping parameter $\kappa$, for the lattices with $N_s=12$ (circles),
$N_s=16$ (triangles), $N_s=24$ (squares), $N_s=32$ (diamonds).
 The full marks are the simulation points, while
the empty ones are extrapolations via the SDM. } 
\label{BIN_CURV}
\end{figure}

As explained in the previous section, for each pair ($N_t,\lambda$)
we have used $N_s=12,16,24,32$. As the field is not compact, it is difficult
to obtain the critical points from observables like the specific heat and 
the susceptibility. An estimate of the critical point for each value of $N_s$
could be obtained from the maximum of the derivative of the 
cumulant $U_{N_s}(\kappa)$. However, in order to minimize
finite size effects, we have rather used the intersections 
$\kappa^*(N_{s_1},N_{s_2})$ 
among all possible pairs of curves $U_{N_s}(\kappa)$, because the differences
between them come only from the corrections to scaling.
After simulating one value of $\kappa$ for every
$N_s$ (having fixed $N_t$ and $\lambda$), we have computed 
$U_{N_s}(\kappa)$ in a narrow interval around 
the simulation point by means of the SDM. 
The simulation point was known to be near the critical
line thanks to some hysteresis or from the estimations for the lattices 
simulated previously. Figure
\ref{BIN_CURV} 
shows $U_{N_s}(\kappa)$ for  
the different values of $N_s$, with $N_t=2$ and $\lambda=0.003$, as an example.
The full marks correspond to the simulation points, while the empty
ones are the extrapolated values.

\begin{figure}[t!]
\epsfig{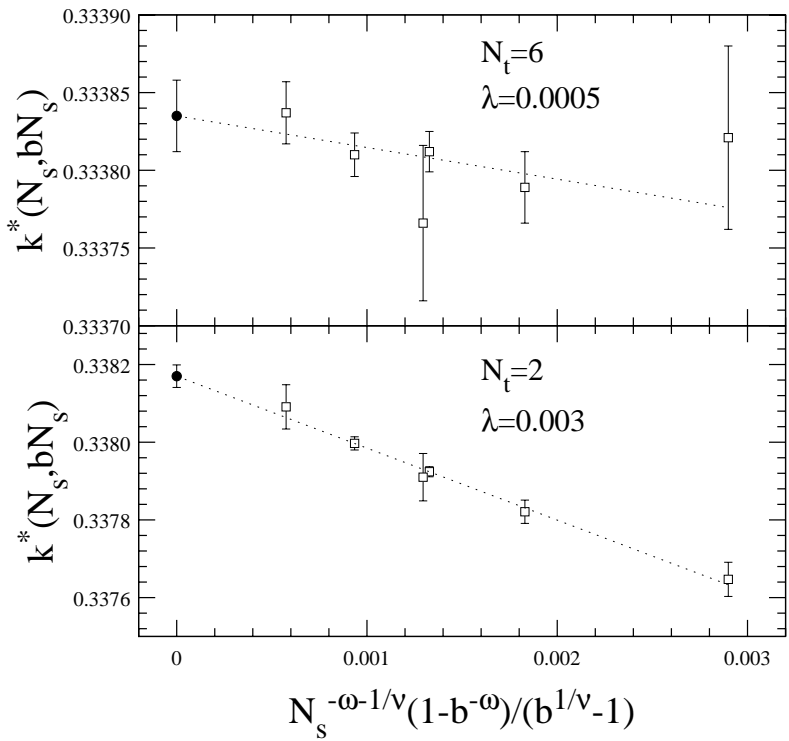}
\caption{ Intersections of Binder cumulants $\kappa^*(N_s, b N_s)$ as
a function of $\frac{1 - b^{-\omega}}{b^{1/\nu}-1}N_s^{-\omega 
-1/\nu}$ (squares). The dotted lines are the fits according to eq.
(\ref{bin}). The full circles are the extrapolated values for
$\kappa_c$ in the limit $N_s \rightarrow \infty$.}
\label{BIN_FIT}
\end{figure}

Starting from the intersections $\kappa^*(N_{s_1},N_{s_2})$, 
the critical 
value $\kappa_c$ in the thermodynamic limit $N_s\to\infty$ 
can be estimated by using the
following law \cite{binder}

\begin{equation}
\kappa^*(N_s,bN_s)-\kappa_c = 
\frac{1-b^{-\omega}}{b^{1/\nu}-1}{N_s}^{-\omega-1/\nu} \label{bin}
\end{equation}
where $\omega$ is the exponent for the corrections-to-scaling. 
Since the scaling parameter is $N_s$, while $N_t$ is fixed, 
according to the hypothesis of dimensional
reduction and universality,
we have used the exponents of the Ising model in two dimensions, 
namely $\nu=1$, $\omega\simeq4/3$ 
\cite{zinn}, obtaining satisfactory fits. To give an idea of
their quality,
in Figure \ref{BIN_FIT} 
we show two extreme
cases, smallest $\lambda$ - largest $N_t$ and 
largest $\lambda$ - smallest $N_t$, 
which are among the worst and  the best fits respectively.

\begin{figure}[t!]
\epsfig{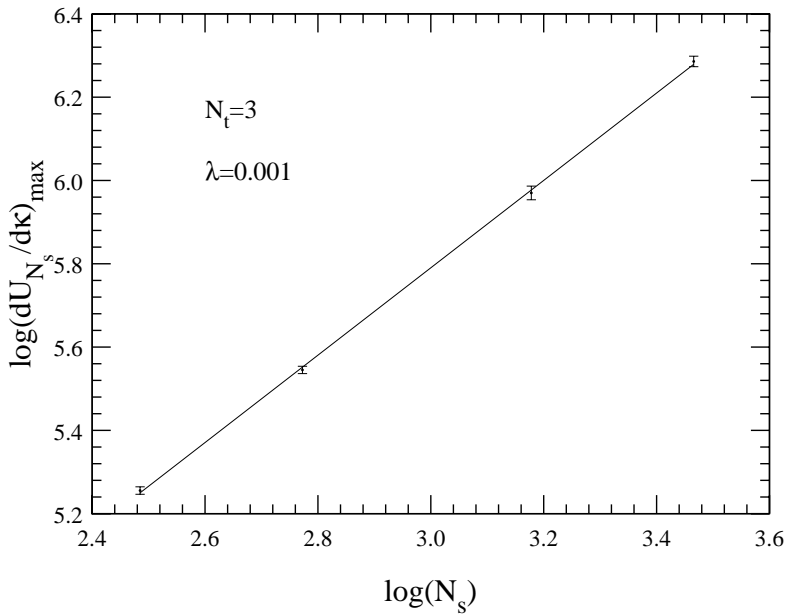}
\caption{Logarithms of the maximal Binder cumulant derivatives
as a function of $N_s$ for $N_t=3$, $\lambda=0.001$. 
The line is the fit to compute $\nu$.}
\label{NU}
\end{figure}

On the other hand,
as a further justification for the use of the two-dimensional exponents, 
we have studied the scaling of the maximum derivative of the
Binder cumulant with $N_s$. It 
was expected to scale like ${N_s}^{1/\nu}$ and we have checked this 
behavior for every pair 
($N_t,\lambda$), obtaining a good scaling and values of $\nu$ 
close to $1$ in every case ($\nu$ ranges
from $0.95$ to $0.99$, and is in all cases almost compatible with $1$ 
within the errors). Figure \ref{NU} 
shows such a fit for
$N_t=3$, $\lambda=0.001$.

\begin{figure}[t!]
\epsfig{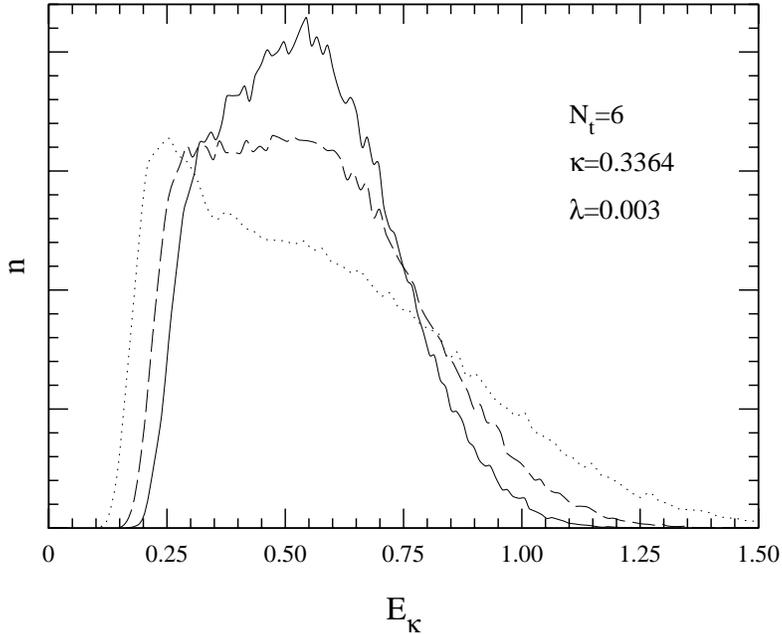}
\caption{ Hystogram of the configurations distribution as a function 
of the term in the action eq.(\ref{fin}) proportional to the hopping parameter 
$\kappa$, for $N_t=6$, $\lambda=0.003$ and $\kappa=0.3364$, 
$N_s=16$ (dots), $24$ (dashes), $32$ (solid).}
\label{HIST}
\end{figure}

Although we have not made an exhaustive study of the order of the transitions,
all the indications point towards the expected second order. One element
is the value of the exponent $\nu$ obtained. 
Another is that no clear discontinuity is 
observed in the histograms for the lattices that we have simulated. 
In Figure \ref{HIST} 
we plot the histograms of the $\kappa$-energy 
for $N_t=6$, $N_s=16,24,32$
at the point $\lambda=0.003$, $\kappa=0.3364$, which is 
very close to the transition line.

\begin{table}[b!]
\begin{center}
\begin{tabular}{|c|c|c|c|c|}
\hline
$\lambda$ & $\kappa_c(N_t=6)$ & $\kappa_c(N_t=4)$  
& $\kappa_c(N_t=3)$ & $\kappa_c(N_t=2)$ \\
\hline
\cline{1-5}
0.003 & 0.336211(30) & 0.336648(17) & 0.337024(30) & 0.338170(29)\\
\hline
0.002 & 0.335249(19) & 0.335567(27) & 0.336007(27) & 0.336711(24)\\
\hline
0.001 & 0.334373(9) & 0.334597(21) & 0.334751(26) & 0.335243(45)\\
\hline
0.0005 & 0.333835(23) & 0.333924(19) & 0.334024(35) & 0.334237(16)\\
\hline
\end{tabular}
\end{center}
\caption{Critical values $\kappa_c$ 
for the different values of $N_t$ and $\lambda$.}
\label{TABLE_KAPPA_C}
\end{table}

\begin{figure}[t!]
\epsfig{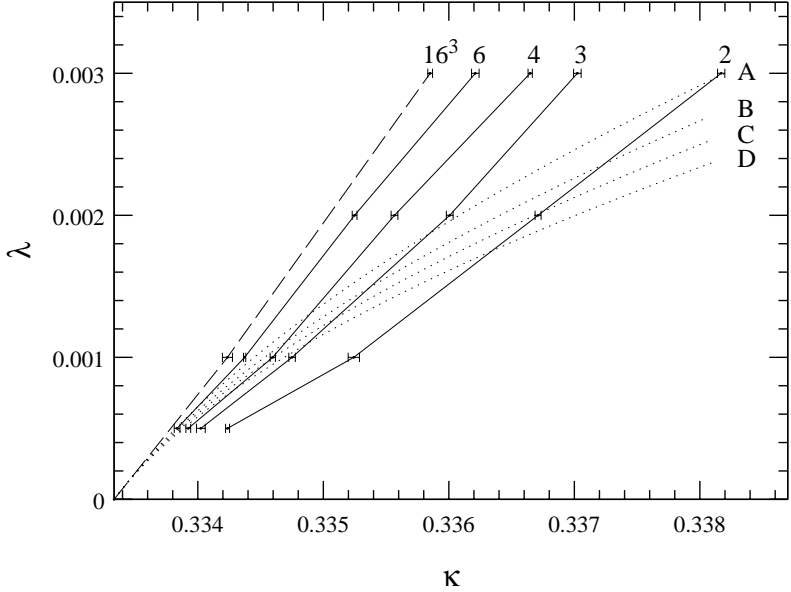}
\caption{ Critical lines (solid curves) for $N_t=2,3,4,6$. 
The dashed line is the 'critical' line for the
symmetric lattice $16^3$. The dotted lines are the
CCP's $A, B, C, D$. }
\label{PH_DIA}
\end{figure}

In table \ref{TABLE_KAPPA_C} 
we quote the values of 
$\kappa_c$ for the different
($N_t,\lambda$)  studied.
In order to have a continuous transition line for every value 
of $N_t$, we have made 
a linear extrapolation between each pair of correlative $\lambda$ values.
Adding the critical line computed for the symmetric lattice 
(estimated on a $16^3$ lattice at $\lambda=0.001,0.003$ and then continued
up to the gaussian point) we have obtained the phase diagram shown in 
Figure \ref{PH_DIA},
where we have also included the CCP's we have used
for the determination of $T^c$.

\subsection{Critical temperatures}

Now we have to search for the intersections of some CCP's with the transition
lines for the different values of $N_t$. 
The region of the phase diagram that we have computed limits the window of 
accessible CCP's, as we could only use CCP's whose intersections with
the critical lines are in the range $0.0005<\lambda<0.003$. In addition, 
in order
to ensure that we were exploring a region of continuum physics, we have
required a good scaling of the correlation length along the trajectories.
Trajectories below $D$ in Figure
\ref{PH_DIA} 
intersect only the critical lines with $N_t=2, 3$ 
and then have been discarded. Trajectories above $A$ have shown 
a poor scaling and have also been eliminated. This could depend
on two circumstances:
on one side the scaling region might be too near the
gaussian point for these trajectories, on the other
it is quite possible that near the critical line the finite size
effects in the measurement of the correlation length are important.

These considerations have led us to consider only the trajectories 
included in the region of the phase diagram 
bounded by the curves $A$ and $D$ in Figure 
\ref{PH_DIA}.
All in all, we have used four trajectories, $A$, $B$, $C$ and $D$,
starting from the points with $\lambda=0.003$ and 
$\kappa=0.3382$, $0.3389$, $0.3395$, $0.3402$ respectively. 
Among them, the only one cutting the critical line for $N_t=6$
is $A$, but it does it almost
tangentially and the error in the determination of the intersection point
is consequently very large.
 This produces an error still larger in the estimate of
the temperature because in the proximity of the gaussian point the
lattice spacing $a$ changes very rapidly along the CCP. As a matter
of fact we have not 
been able to use $N_t=6$ for the measurement of the critical temperature 
and we had to limit ourselves to the crossings of the CCP's 
with the critical lines for $N_t=2, 3, 4$.

\begin{table}[t!]
\small
\begin{center}
\begin{tabular}{|c|c|c|c|c|c|c|}
\hline
$N_t$ & $\kappa$ & $\lambda$ & $\xi$ & $\xi \cdot s$  
&  $\frac{-16 {\pi}^2 m^2(T_{N_t})}{g^2}$ & $\frac{4 \pi T_{N_t}}{g}$ \\
\hline
\cline{1-7}
2 & 0.33810(8) & 0.00296(3) & 3.42(3) & 3.38(3) & 17.275(2) & 10.11(11)\\
\hline
3 & 0.33584(9) & 0.00187(5) & 5.78(9) & 3.65(6) & 17.268(4) & 10.55(28)\\
\hline
4 & 0.33489(7) & 0.00130(4) & 8.84(12) & 3.91(6) & 17.256(5) & 11.29(35)\\
\hline
\cline{1-7}
2 & 0.33722(6) & 0.00236(3) & 3.75(5) & 2.98(4) & 22.389(2) & 12.64(14)\\
\hline
3 & 0.33536(8) & 0.00148(5) & 7.10(14)& 3.59(7) & 22.381(5) & 13.23(38)\\
\hline
4 & 0.33469(6) & 0.00110(3) & 9.36(16) & 3.53(6) & 22.381(5) & 13.29(39)\\
\hline
\cline{1-7}
2 & 0.33674(6) & 0.00202(3) & 3.94(8) & 2.70(6) & 26.829(2) & 14.68(17)\\
\hline
3 & 0.33508(8) & 0.00127(4) & 7.27(32)& 3.15(15) & 26.821(5) & 15.45(48)\\
\hline
4 & 0.33456(9) & 0.00097(5) & 9.92(28) & 3.31(10) & 26.825(9) & 15.06(86)\\
\hline
\cline{1-7}
2 & 0.33634(7) & 0.00175(3) & 4.11(8) & 2.45(5) & 32.074(3) & 16.95(26)\\
\hline
3 & 0.33486(7) & 0.00109(4) & 8.02(25)& 3.02(10) & 32.065(6) & 17.91(60)\\
\hline
4 & 0.33427(12) & 0.00076(7) & 10.93(19) & 2.86(6) & 32.052(16) & 19.28(1.89)\\
\hline
\end{tabular}
\end{center}
\caption{ Correlation lengths $\xi$ and estimates $T_{N_t}$ of the
critical temperatures associated with the crossings of the CCP's
$A, B, C, D$ (from top to bottom) with the critical lines of the
lattices with $ N_t= 2, 3, 4 $.}
\label{TABLE_TG}
\end{table}

 In table \ref{TABLE_TG} we show 
the values of $\kappa$ and $\lambda$ at the points where the trajectories 
$A$, $B$, $C$ and $D$ 
(from up to down) intersect the critical lines for $N_t=2, 3, 4$.
Their errors come from the deviations in the estimation of the critical lines, 
and have been calculated as explained below for the critical temperatures. 

In correspondence with these intersection points 
we have measured the correlation
length on a $16^3$ lattice in order to verify its scaling. 
We have chosen these and not any other points on the CCP's because the good or 
poor scaling of $\xi$ just there indicates which values of $N_t$
we will be able to use for estimating the critical temperature in the 
continuum. The values of $\xi$ obtained are quoted in table \ref{TABLE_TG}.
The errors quoted are the ones arising from the simulation and 
have been calculated with the jackknife method.  

When going towards the gaussian point,
$\xi$ should grow inversely to the parameter $s$, i.e. inversely to $a$
(see. eq.(\ref{scal})) and thus the product $\xi\cdot s$ should be constant
along the trajectory
(the absolute scale of $s$ is of course arbitrary and, for each CCP,
we have taken $s=1$ at its starting point).
It is apparent from table \ref{TABLE_TG} that for
the trajectories $B$, $C$ and $D$ the values of $\xi \cdot s$ for $N_t=3,4$
are fully compatible within the errors, while for $N_t=2$ there are 
significant deviations from scaling. For the trajectory 
$A$ the result is less satisfactory, something that will
be reflected also in a small difference among the estimates for the
critical temperature 
resulting from $N_t=3$ and $4$ for this trajectory.

As we said above, we have constructed the critical lines
by means of a linear interpolation of the critical points 
obtained numerically and have then used them to estimate the 
intersections with the CCP's. From Figure 
\ref{PH_DIA} 
it is apparent
that this is a rather good approximation. We have estimated the errors in
the measurement of $T_{N_t}
/g$ and ${m^2_{MS}} / g^2$ in the following
way. In addition to the line formed by joining the
values of $\kappa_c(\lambda)$ for the four values of $\lambda$ simulated,
we have considered two more lines, on the two sides of the
previous one. One has been 
obtained by joining the points corresponding to the upper values
$\kappa_c(\lambda)+\sigma(\kappa_c(\lambda))$ 
and the other by joining the points corresponding to the lower values
$\kappa_c(\lambda)-\sigma(\kappa_c(\lambda))$, 
$\sigma(\kappa_c(\lambda))$ being the error in the determination of 
$\kappa_c(\lambda)$. Each of the three lines gives a different 
estimate for the
observables. We have taken as our estimate the one given by the 
central line and the error
has been taken to be one half of the difference between the 
values given by the two external lines. This procedure 
would overestimate the
error if the critical lines between the critical points computed 
were really straight; 
this overestimation should approximately compensate for the error 
introduced by the linearization. 

The values of $T_{N_t}/g$ and ${m^2_{MS}} / g^2$ 
obtained in this way are shown in table 
\ref{TABLE_TG}
(multiplied by the constant factors $4\pi$ and $-16\pi^2$ in order to compare 
directly with the results in \cite{ein}). 
It can be seen how, for $N_t=3,4$, $T_{N_t}/g$ 
has reached a well defined limit (only for the
trajectory $A$ the values of 
$T_{3}/g$ and $T_{4}/g$ 
are not perfectly compatible within the errors). 

\begin{figure}[b!]
\epsfig{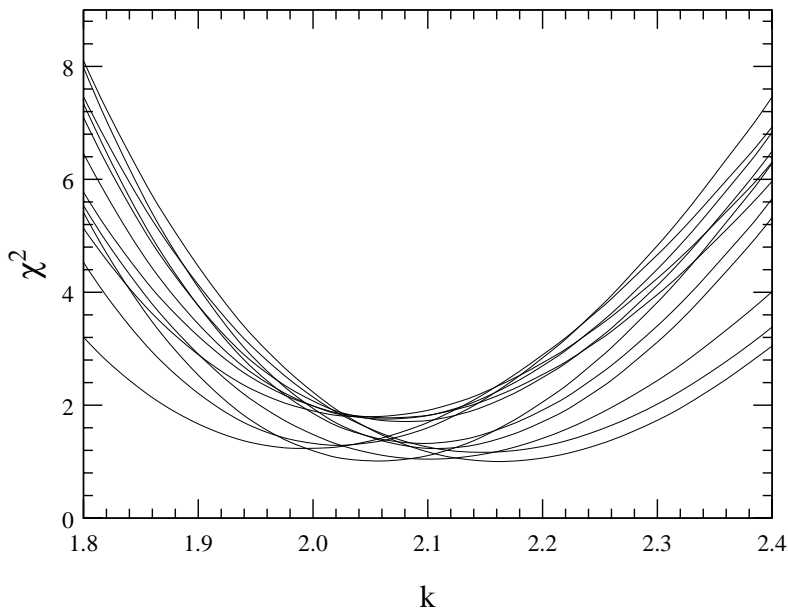}
\caption{Plots of $\chi^2$ as a function of the parameter $k$ 
in eq.(\ref{ein})
for all maximal subsets of statistically independent data extracted from 
table 
\ref{TABLE_TG} 
(the point relative to the CCP $A$ and $N_t=3$ has been 
excluded, so as all those of $N_t=2$). The number of degrees of freedom in 
all cases is $5-1=4$.
}
\label{CHI2}
\end{figure}

In order to
check eq.(\ref{ein}), we have performed a series of best fits. 
To this purpose we have used all the data relative to
$N_t=3,4$, except the point corresponding to the
intercept of trajectory $A$ with the critical line for $N_t=3$,
for the reason explained above. Some care has been necessary,
because not all the data are statistically independent.
Consider for instance the intercepts of trajectories
$B, C, D$ with the critical line for $N_t=3$: they belong to the
same straight segment of the interpolated critical line, which implies that
among them only two are statistically independent of each other, while the
third can be determined thereof. For this reason, we could not use
all the data at once in a single fit; rather we have 
repeated the fits for
all possible maximal subsets of statistically independent points extracted
from the data. It turned out that all subsets consisted of five
points out of a total of seven. The plots of 
$\chi^2(k)$ for all the combinations of data are displayed
in Figure \ref{CHI2} (notice that $\chi^2(k)$ is not normalised
per degree of freedom). 

\begin{figure}[t!]
\epsfig{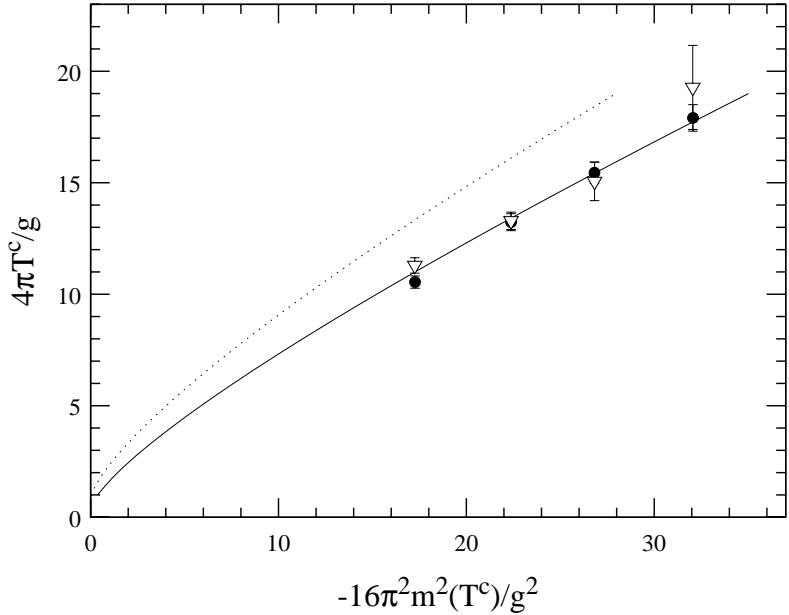}
\caption{ Estimates $T_{N_t}$ of the critical temperature 
as a function of the renormalised parameters $m^2_{MS}(T_{N_t})$ and $g$, 
for the CCP's $A, B, C, D$ (from left to right), $N_t=3$ (full
circles) and $N_t=4$ (empty triangles). The solid line is the 
fit according to eq.(\ref{ein}), with $k=2.1$. The dotted line 
corresponds to $k=1$.}
\label{TG_MG}
\end{figure}

As can be seen from the figure, the errors on the estimates of
$k$ (equal to semi-widths of the intervals  
in $k$, around the minimum, for which
the variation of $\chi^2$ is less than $1$) 
are comparable for all
the combinations of data and are around $0.15$. 
It is apparent that in all cases the minimum value
of $\chi^2(k)$ per degree of freedom (four) is very good, which
indicates that the fits can be trusted. The values of $k$ obtained
from the fits are all practically consistent with each other, within
the errors. As our final estimate, we have taken their average,
while for the error we have taken a value larger than the
individual errors, in such a way that all the
fits are compatible with it. In conclusion, we have got: $k=2.1(2)$. 
The final fit is shown in Figure \ref{TG_MG},where the curve having
$k=1$ is also displayed for comparison.

\section{Conclusions}

We have investigated the phenomenon of symmetry restoration
at finite temperature in the scalar $\lambda\phi^4_3$ model,
by means of a Monte Carlo simulation. Using FSS techniques
we have carefully determined the critical lines separating the ordered
and the unordered phases on asymmetric lattices, for a
number of different extensions in the temporal direction.
The intersections of these lines with various RG trajectories 
lying in the ordered phase of the $T=0$ theory 
provide estimates of the critical temperature.
The analysis of the results for lattices
with increasing extensions in the time-direction clearly shows
that these temperatures approach a well-defined finite value
in the continuum limit, confirming the expectation of a symmetry
restoring phase transition at a finite physical temperature for 
the continuum theory. We have measured $T^c$ for various RG
trajectories.
In this way we have checked the validity of the following
equation relating $T^c$ to the renormalized parameters
of the continuum theory:
$$
T^c \log \left( \frac{4 \pi k T^c}{g} \right)
= - \frac{8 \pi m^2_{MS}
(T^c)}{g}~,
$$
which was derived in \cite{ein} using renormalization group
methods, and have been able to measure the so far unknown parameter
$k$, obtaining the value $k=2.1(2)$.

\section{Acknowledgments}

The authors are grateful to J.L. Alonso, M. Asorey, J.M. Carmona and L.A.
Fern\'andez for discussions while
the manuscript was in preparation.
This work is
partially supported by CICyT AEN94-0218, and
AEN95-1284-E. D. I\~niguez and C.L. Ullod are MEC and DGA fellows
respectivelly.

\end{document}